\documentclass{ar-1col}
\usepackage{url}
\usepackage[numbers]{natbib}
\usepackage{amsmath}
\usepackage{amssymb}
\usepackage{graphicx}
\jname{Xxxx. Xxx. Xxx. Xxx.}
\jvol{AA}
\jyear{YYYY}
\doi{10.1146/((please add article doi))}

\begin{document}

\markboth{Paulsen and Keim}{Mechanical memories in solids}

\title{Mechanical memories in solids, from disorder to design}

\author{Joseph D.\ Paulsen$^1$ and Nathan C.\ Keim$^2$
\affil{$^1$Department of Physics and BioInspired Institute, Syracuse University, Syracuse, NY, USA 13244; email: jdpaulse@syr.edu}
\affil{$^2$Department of Physics, Pennsylvania State University, University Park, PA, USA 16802; email: keim@psu.edu}}

\begin{abstract}
Solids are rigid, which means that when left undisturbed, their structures are nearly static. It follows that these structures depend on history---but it is surprising that they hold readable memories of past events. 
Here we review the research that has recently flourished around mechanical memory formation, beginning with amorphous solids' various memories of deformation and mesoscopic models based on particle rearrangements. 
We describe how these concepts apply to a much wider range of solids and glassy matter---and how they are a bridge to memory and physical computing in mechanical metamaterials. 
An understanding of memory in all these solids can potentially be the basis for designing or training functionality into materials. 
Just as important is memory's value for understanding matter whenever it is complex, frustrated, and out of equilibrium. 
\end{abstract}

\maketitle

\section{INTRODUCTION}

Suppose a friend asks you a simple “yes or no” question (like, “do you want to go out to dinner tonight?”), and then inexplicably hands you a bucket of sand. Before your friend leaves the room, they instruct you to encode your answer in the sand in the following manner: clockwise for “yes”, counterclockwise for “no”. You are left somewhat puzzled, but after some consideration you realize that by inserting a solid rod in the center of the bucket and twisting, you can indeed encode a single bit of memory in this disordered pile of grains.

\begin{figure}[b]
\includegraphics[width=4.9in]{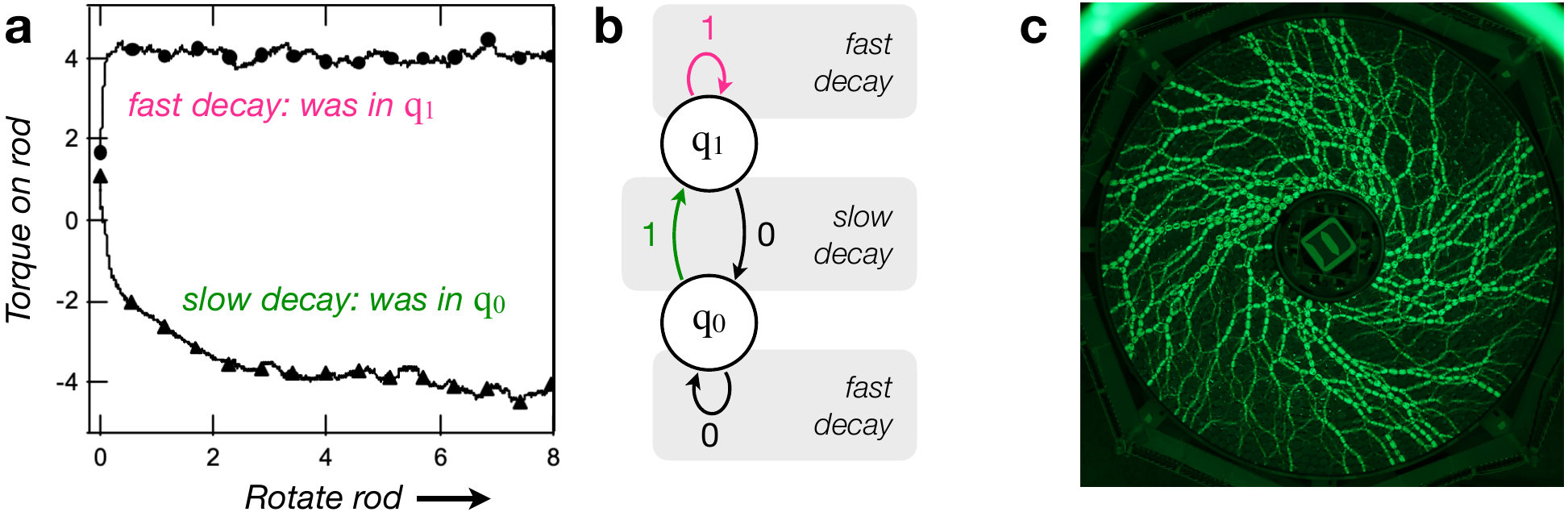}
\caption{
Memory of a shear direction. 
\textbf{(a)} Mechanical response of a granular material during clockwise shear between two concentric cylinders~\cite{Toiya04}. $q_0$ and $q_1$ are states after counterclockwise and clockwise rotation, respectively.
If the operator resumes shearing in the same direction that was applied last, the material shows a fast decay to a steady state response. 
If the shear direction is reversed, the decay is slower. 
\textbf{(b)} Corresponding graph of states and transitions, representing encoding and readout of a single bit: ``1'' edges represent the clockwise rotations in (a); ``0'' edges are counterclockwise. 
\textbf{(c)} Monolayer of photoelastic discs, revealing the anisotropic load-bearing network that encodes information in this system~\cite{Zhao19}. 
Panel \emph{a} adapted with permission from Ref.~\cite{Toiya04}: Toiya, Stambaugh, and Losert, \emph{Phys.\ Rev.\ Lett.}\ 93, 088001 (2004). Copyright 2004 by the American Physical Society.
Panel \emph{c} courtesy of Yiqiu Zhao.
}
\label{fig:1}
\end{figure}

The key is a mechanical response of a granular material that was demonstrated by Toiya \textit{et al.}~\cite{Toiya04}. 
To encode a memory in the material, you slowly rotate the rod about its vertical axis in your chosen direction (clockwise or counterclockwise) and then stop the rotation, leaving the rod in place. 
To read out the memory, your friend rotates the rod in one direction (clockwise, say). If they don't see a significant transient (top curve in Fig.~\ref{fig:1}a), they know you too rotated the rod clockwise. If they see a transient (bottom curve), they know you rotated the rod counterclockwise, i.e., opposite their “readout”. The entire process of writing and readout is captured by Fig.~\ref{fig:1}b.

What can we learn from this memory game? It is not surprising that collections of grains display history dependence. The surprise is the clean signature that comes from a simple macroscopic protocol---shearing the material from its boundaries. This strong, unambiguous response of the system that reveals aspects of its past can be seen as a “memory” formed by ordinary matter. 
Understanding what a material can remember, and what microscopic mechanisms underlie these abilities, are topics of intense contemporary interest~\cite{Keim19,Barrat24}. 

In this article we first examine memory in amorphous solids, which will serve as a foundation from which we can generalize to other solids, and which can even form a useful bridge to memories in mechanical metamaterials. While shape memory---most notably the recovery of a shape upon heating \cite{Lagoudas08,Mather09,Keim19}---has important connections with the examples here, it operates by distinct mechanisms, and so our brief review omits it. Remarkably, the wealth of memories to come is revealed by deformations alone.

\subsection{Connecting memory and material}

Returning to the binary digit in the bucket of sand:
how is the memory of direction connected with the physics of a granular packing?
Figure~\ref{fig:1}c shows a layer of discs that accomplish the same feat \cite{Howell99,Zhao19}. 
The discs are photoelastic so that when they are viewed through cross-polarizing filters, they reveal a network of load-bearing contacts within the material \cite{Zadeh19}. 
As the outer boundary is rotated clockwise, the grains form an anisotropic state \cite{Majmudar05} with chains of strong contacts
spiraling between the outer and inner walls. 
If the rotation is stopped and restarted in the same direction, this anisotropy remains and a steady state is quickly resumed, corresponding to the top curve in Fig.~\ref{fig:1}a. If the rotation proceeds in the opposite direction, the particles must negotiate a transition to a different anisotropic structure. This process requires a finite rotation angle to complete before settling into a steady state; it is the source of the transient in the bottom curve in Fig.~\ref{fig:1}a. A qualitatively similar response is also observed in concentrated suspensions of solid spheres~\cite{GadalaMaria80}. 
This example sets the pattern for the discussions that follow: our understanding of a material's memory is tied to our understanding of its physics.

\subsection{Memory of an amplitude}

If you had never considered the idea of storing information in a disordered pile of sand, the previous example may come as somewhat of a surprise. 
Nevertheless, the underlying mechanism is somewhat intuitive, arising from the way microstructure imparts rigidity.
The next example, however, stretches our imaginations for how and what a material can learn. 

\begin{figure}[tb]
\includegraphics[width=6.3in]{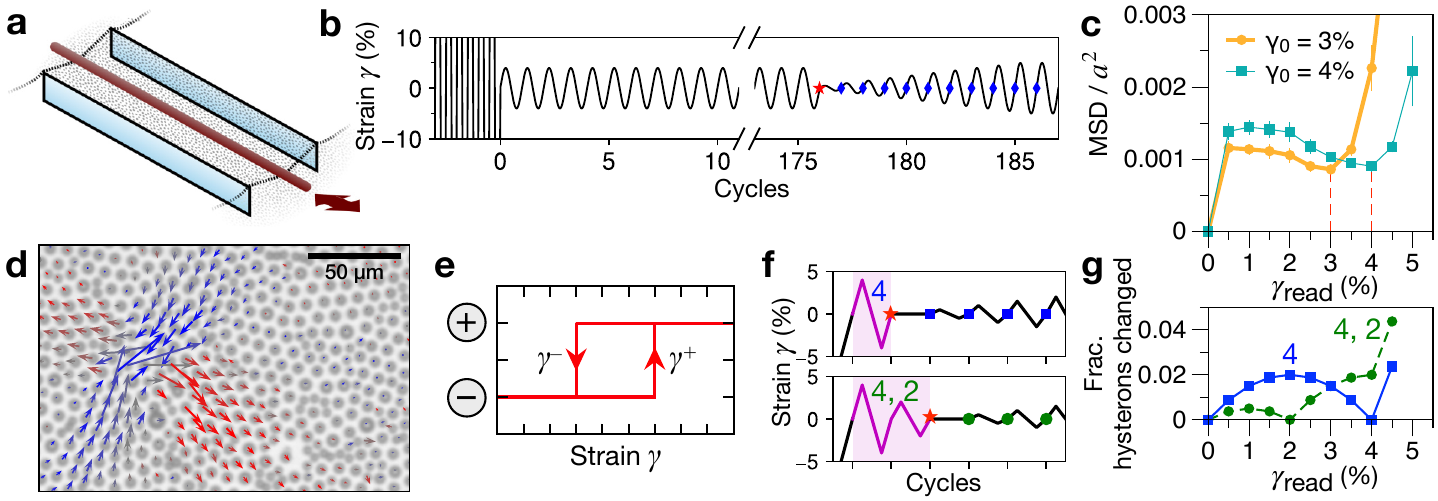}
\caption{
    Memories of amplitude.
    \textbf{(a)} Schematic of a 2D amorphous solid experiment~\cite{Keim20}. Colloidal particles with electrostatic repulsion are adsorbed at an oil-water interface. A magnetic needle at the interface (red) moves within a narrow channel to shear the material. 
    \textbf{(b)} Shear protocol for experiment: ``reset'' with large amplitude $\sim$50\%, then 176 cycles with strain amplitude $\gamma_0 = 4\%$, and memory readout with cycles of increasing amplitude $\gamma_\text{read}$. After each readout cycle (blue diamonds), particle positions are compared to start of readout (red star). 
    \textbf{(c)} Mean squared particle displacements during readout, normalized with particle spacing $a$. After preparation with $\gamma_0 = 3\%$ or $4\%$, minimum at $\gamma_\text{read} = \gamma_0$ indicates memory. 
    \textbf{(d)} Image from monolayer experiment. Superposed arrows are displacements magnified 10$\times$ from a single plastic rearrangement event, colored by direction. 
    \textbf{(e)} Hysteron model for rearrangement switching between ``$+$'' and ``$-$'' state at strain thresholds $\gamma^+$ and $\gamma^-$.
    \textbf{(f)} Strain protocols for an ensemble of hysterons (Preisach model). Top: Writing and reading memory of 4\%. Bottom: Two cycles write 4\% and 2\%.     \textbf{(g)} Fraction of hysterons in Preisach model that differ from start of readout, for protocols in (f).
    Panel \emph{a} reproduced from Ref.~\citealp{Keim20} (CC BY 4.0).
    Panels \emph{b--d} adapted from Ref.~\citealp{Keim22} (CC BY 4.0).
}
\label{fig:amplitude}
\end{figure}

Figure~\ref{fig:amplitude}a illustrates an experiment where a solid made of particles at an oil-water interface is cyclically sheared by a steel needle adsorbed at the same interface \cite{Keim20}. This material is a model of amorphous solids generally, with the advantage that $\sim$40,000 particles can be imaged and tracked to study the relationship between deformation and structure.
In the protocol shown in Fig.~\ref{fig:amplitude}b, the sample is sheared sinusoidally with a fixed strain amplitude $\gamma_0$ for 176 cycles. After this training period, the experimenter applies a series of shear cycles with increasing amplitude $\gamma_\text{read}$. Figure~\ref{fig:amplitude}c measures the rearrangements that are observed with respect to the  state at the end of training, as a function of the applied readout strain amplitude. 

Intriguingly, the material has “learned” the strain value from the training period. The state reached by training with amplitude $\gamma_0 = 3\%$ is almost completely recovered by shearing again at $\gamma_0 = 3\%$---while it is disrupted by the much weaker strain of $\gamma_\text{read} = 0.5\%$. Likewise, the system trained at $\gamma_0 = 4\%$ exhibits the fewest net rearrangements at a readout strain of $\gamma_0 = 4\%$.
The molecular dynamics simulations of Fiocco \emph{et al.\ }were the first to see this behavior, plus an even greater surprise: 
when training at 4\% is followed by a final cycle of amplitude 3\%, both memories are evident in the readout
\cite{Fiocco14,Mukherji19,Keim20}. In Secs.~\ref{sec:rpm} and \ref{sec:annealcap} we will explore this multiple-memory capacity in models and experiments. 

\begin{marginnote}[]
\entry{rearrangement}{A localized relaxation event that changes particles' relative positions and dissipates energy.}
\entry{soft spot}{A localized region in an amorphous solid that is mechanically weaker and more likely to rearrange under load.}
\entry{reversible plasticity}{Plastic rearrangements of microstructure in response to a deformation, that can be perfectly undone by another deformation.}
\end{marginnote}

\subsection{Reversible plasticity and return-point memory}
\label{sec:rpm}

Reflecting on the above experiment, one may be shocked that there is any orderly response at all, from a jumble of particles that are endlessly moving around one another. That the paths of particles are anything but a chaotic mess is stunning, let alone that they encode some amount of meaningful information. 
Recent work has traced this orderly response to localized groups of particles that collectively rearrange under loading (Fig.~\ref{fig:amplitude}d). 
Notably, many of these rearrangements are undone when the loading is reversed, so that each particle returns to its previous position. 
This process is hysteretic as schematized in Fig.~\ref{fig:amplitude}e. 
Under cyclic loading, the plastic deformation of an entire amorphous solid with many such regions can be perfectly reversed by the end of each cycle---a response termed ``reversible plasticity''.

These behaviors bring to mind the classic Preisach model~\cite{Preisach35,Semenov2024} for 
hysteresis in response to an imposed field or other driving parameter---here, the strain $\gamma$. The model considers many elementary units or ``hysterons,'' each with just two possible states, ``$+$'' and ``$-$''. The $i$th hysteron switches to each state when $\gamma \ge \gamma_i^+$ or $\gamma \le \gamma_i^-$, respectively (Fig.~\ref{fig:amplitude}e). 
While in many applications the hysterons are notional and their thresholds are fitting parameters~\cite{Magergoyz1985,Guyer1999,Guyer2006}, observations show that in this case, each rearranging region acts as a hysteron~\cite{Lundberg08,Keim14,Mungan19,Keim20}. Indeed, by measuring each rearrangement's $\gamma^\pm$ in a cycle at one strain amplitude, one can construct a Preisach model that predicts the system's response at smaller amplitudes~\cite{Keim22}. Hysterons and their collective behaviors are a starting point for understanding the wealth of behaviors to come in this review.

\begin{marginnote}[]
\entry{hysteron}{A basic element of hysteresis, used to model bistable subsystems such as reversible particle rearrangements at soft spots.} \entry{Preisach model}{A group of non-interacting hysterons responding to an imposed driving field, e.g.\ shear strain.}
\end{marginnote}

One crucial property of the Preisach model is that since the thresholds $\gamma_i^\pm$ never change, there is a fixed sequence in which transitions happen as $\gamma$ is increased, and another fixed sequence as $\gamma$ is decreased. A consequence of this property is that whenever the driving is bounded between two values $\gamma_\text{min}$ and $\gamma_\text{max}$, the system's evolution is likewise bounded~\cite{Sethna93}---a rule that resembles the ``no-passing'' property of sliding charge-density waves~\cite{Middleton92}. For example, if the $i$th hysteron is observed in the ``$-$'' state when $\gamma$ is at both of these bounds, then it can never be in the ``$+$'' state as long as $\gamma$ stays bounded. This property implies that as long as $\gamma_\text{min} \le \gamma \le \gamma_\text{max}$, revisiting either bound will put \emph{all} hysterons into the same state as when that bound was last visited~\cite{Sethna93}---a behavior called return-point memory~\cite{Barker83, Sethna93,Keim19,Terzi20}. One can simulate an ensemble of hysterons with random $\gamma^\pm$, and drive it with protocols in Fig.~\ref{fig:amplitude}f that mimic experiments, measuring the fraction of hysterons that differ during readout. The results in Fig.~\ref{fig:amplitude}g qualitatively match experiments, and they are consistent with return-point memory: the training cycles established bounds at $\pm \gamma_0$, and even though the initial readout cycles with $\gamma_\text{read} < \gamma_0$ change the system, the cycle with $\gamma_\text{read} = \gamma_0$ places the system back onto the same trajectory as before, restoring its state.

Return-point behavior is also recursive: for a portion of the time that strain is bounded between $[\gamma_\text{min}, \gamma_\text{max}]$, it can be further bounded by $[\gamma'_\text{min}, \gamma'_\text{max}]$ nested within that outer pair, and so on~\cite{Barker83}. Thus for cyclic driving, the state of the system at $\gamma = 0$ depends on the history of \emph{nested} turning points. As shown in Fig.~\ref{fig:amplitude}(f, g), applying the sequence of two strain amplitudes $\gamma_0 = 4\%, 2\%$ will generally yield a different outcome than $2\%$ or $4\%$ alone---whereas the result of $\gamma_0 = 2\%, 4\%$ is indistinguishable from 4\%. 
Crucially, to avoid revisiting a turning point and losing the new memory, \emph{both} new bounds must be inside the previous interval. For example, omitting all $\gamma < 0$ in Fig.~\ref{fig:amplitude}f, so that $\gamma'_\text{min} = \gamma_\text{min}$, will fail to write a second memory.

To look for memories of $2\%$ and $4\%$, one sweeps the strain amplitude (or, in some cases, strain) from small to large values, exiting the nested intervals from smallest to largest. 
Leaving an interval means revisiting one of its turning points and erasing the history of smaller deformations since that memory was written. This loss of history changes the system's differential response to further increases in strain, indicating a memory~\cite{Barker83,Guyer1999,Guyer2006,Keim19}. Return-point memory is the same principle that lets a combination lock store multiple values from the back-and-forth rotation of a single knob \cite{Keim19}.
% In Sec.~\ref{sec:annealcap} we will consider how these concepts apply in real materials.

\subsection{Memory capacity and computation}

Now that we have touched on two types of memory that appear to be qualitatively different, we look at another way of seeing that their physics is distinct: namely, their capacity for multiple memories. Memory acts as a filter to preserve only certain features of history: a memory of direction retains only the sign of the most recent deformation, while in return-point memory, multiple amplitudes can be stored only in descending order. We can also estimate the number of bits required to encode the information in readout---i.e.\ the base-2 logarithm of how many distinct histories readout can represent. Here too, differences are apparent: for memory of a direction this measure is independent of system size $N$, while for return-point memory it scales with $\sqrt{N}$ \cite{Regev21,Keim22} and for computer memory it scales as $N$. 
Whether it is probed qualitatively or quantitatively, memory capacity is a tool for describing the non-equilibrium nature of the solids in this review, and of countless other systems~\cite{Keim19}.

\section{MEMORY AND ANNEALING} \label{sec:anneal}

In our description of reversible plasticity and return-point memory, it is evident that many repeated cycles of a deformation have the same net effect as a single cycle. 
Yet an amorphous solid that has been freshly formed or deformed catastrophically (strains of order 100\%) must be sheared for many cycles before it shows reversible plasticity and return-point memory---here, memory capacity itself depends on history. Each cycle tends to make fewer changes than the previous one, until finally, for $\gamma_0$ below some empirical threshold $\gamma_c \sim 10\%$, there are no net changes at all. In contrast, for $\gamma_0>\gamma_c$, changes continue indefinitely~\cite{Regev13,Keim13,Regev15,Royer15,Priezjev13,Fiocco14,Nagamanasa14,Mungan19,Adhikari18,Zhao22,Reichhardt23}. The existence of the critical strain $\gamma_c$ is reminiscent of reversible-irreversible transitions in other memory-forming systems, and in solids it seems connected with the yielding transition by which the material loses rigidity and begins to flow \cite{Reichhardt23,Nagamanasa14,Keim13,Regev13}. 
The transient ``mechanical annealing’’ below $\gamma_c$ resembles its thermal counterpart: a molecular glass held at sufficient temperature will explore different configurations and ultimately reach ones that are more stable and long-lived.

\begin{figure}[t]
    \includegraphics[width=6in]{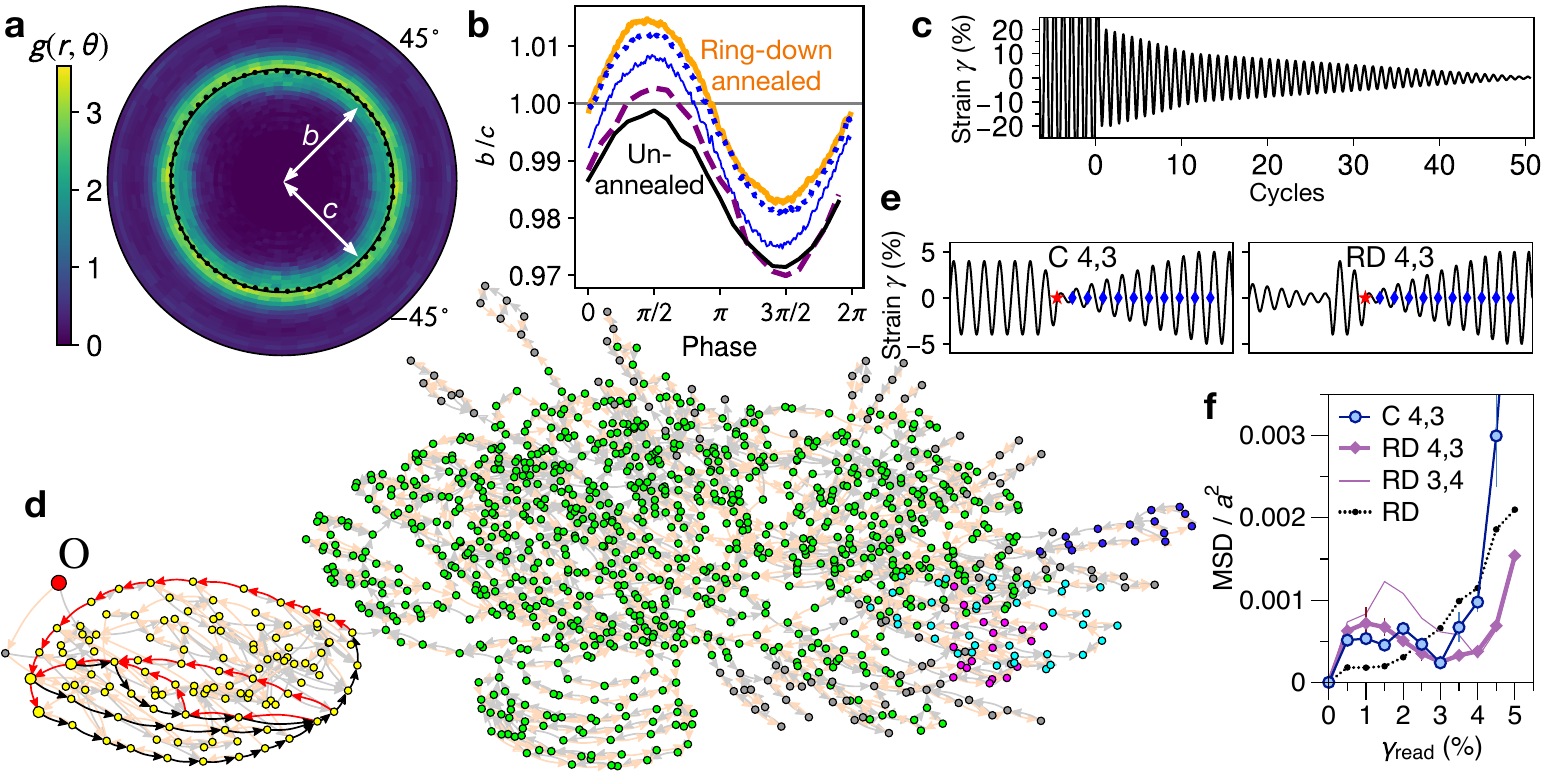}
    \caption{
        Relationships between annealing and memory.
        \textbf{(a)} Memory of direction in an un-annealed experimental sample. Pair correlation function $g(r, \theta)$ represents the average positions of each particle's neighbors. Black dots mark peak at each $\theta$. Comparison with black circle shows elongation along $-45^\circ$. Ellipse fit to the peaks has semiaxes $b$, $c$~\cite{Keim22}. 
        \textbf{(b)} Anisotropy $b/c$ in a cycle with amplitude 3.5\%, after various annealing protocols. Ring-down fully relaxes anisotropy seen in unannealed samples.
        Unlabeled curves: annealed with $\gamma_0 = 3.5\%$ (dashed purple), and two trials with $\gamma_0 = 4\%$ (solid and dotted blue curves).
        \textbf{(c)} Ring-down protocol to relax anisotropy in (b): after reset with amplitude 50\%, material is annealed by slowly decreasing amplitude.
        \textbf{(d)} Excerpt of transition graph by Regev \emph{et al.}~\cite{Regev21} showing states (circles) reachable by forward (black arrows) and reverse (red) shear, starting at un-annealed state $O$. Most arrows are faded for clarity. Like-colored states form strongly-connected components featuring many limit cycles (examples shown with un-faded arrows).
        \textbf{(e)} Protocols for writing and reading multiple memories. Left: writing both memories after ring-down annealing. Right: writing the smaller memory after constant-amplitude annealing.         
        \textbf{(f)} Experimental multiple-memory readouts. Curves corresponding to (e) show effect of annealing envelope at larger $\gamma_\text{read}$. 
        Switching order of memories (``RD 3,4'') erases the smaller one. Ring-down without writing yields ``RD'' curve. 
        Panels \emph{a, b, c, e, f} adapted from Ref.~\citealp{Keim22} (CC BY 4.0).
        Panel \emph{d} adapted with permission from Ref.~\citealp{Regev21}: Regev, Attia, Dahmen, Sastry, and Mungan, \emph{Phys.\ Rev.\ E} 103, 062614 (2021). Copyright 2021 by the American Physical Society.
        }
    \label{fig:anneal}
    \end{figure}

The need for annealing is tied to the non-equilibrium character of solids at rest: even obvious differences from a configurational ground state might not relax. Figure~\ref{fig:anneal}a illustrates this point with the pair correlation function $g(r, \theta)$ that represents the average positions of every particle's neighbors in an experimental sample, after a large cyclic deformation with amplitude $\sim$50\% has erased the effects of past experiments~\cite{Keim13,Keim22}.
Close inspection shows that neighbors are slightly more distant along the $-45^\circ$ direction---one of the principal axes of the preceding shear. We measure this memory of direction as the ratio of ellipse axes $b/c$~\cite{Teich21,Galloway22,Keim22,Edera24}. Figure~\ref{fig:anneal}b shows that this anisotropy persists---not only at rest, but even with cyclic shear at $\gamma_0 = 3.5\%$. Relaxing the average $b/c$ to zero requires larger deformations---either a larger constant value (here, 4\%), ideally just below $\gamma_c$~\cite{Galloway22,Edera24}, or a gradual ``ring-down'' (Fig.~\ref{fig:anneal}c).

Furthering the resemblance to thermal annealing, recent molecular dynamics simulations~\cite{Bhaumik21}, mesoscopic models~\cite{Liu22,Sastry21} and experiments~\cite{Edera24,Zhao22} suggest a partial overlap between mechanical annealing and the better-known thermal relaxation of glasses. 
In this picture, the configurations that are reversible under the largest deformations tend to have the lowest structural energy, and they are the fewest and take the longest to find. 
Much like temperature, increasing strain amplitude allows additional structural changes in a material, expanding the set of accessible configurations while narrowing the criterion for success, until reversible plasticity becomes impossible at $\gamma_c$. Thus, increasing $\gamma_0$ after annealing will always induce further annealing, perhaps for multiple cycles---the annealed system is marginally stable at $\gamma_0$~\cite{Strogatz94}. 
However, mechanical annealing is ultimately at odds with the most thorough ways of relaxing amorphous solids: special ``ultrastable'' solids formed in simulations have structural energies far below what is possible mechanically~\cite{Bhaumik21,Liu22,Sastry21}, and seem incapable of forming memories~\cite{Arceri21}. 

\begin{marginnote}[]
    \entry{transition graph}{A directed graph where each node represents a stable configuration of particles, and each edge is a rearrangement event.}
    \entry{quasistatic deformation}{A deformation performed in many small strain steps, approximating the limit of zero strain rate and zero inertia.}
    \entry{marginal stability}{A stable state obtained at some value or amplitude of driving, that loses stability when subjected to a finite perturbation.}
\end{marginnote}

Inspired by the previous section, we can also begin to see annealing in discrete terms. In a quasistatic simulation, one can distinguish the strain steps that cause rearrangements---signaled by a sudden drop in energy---from the steps in which the material remains essentially in the same state. Each time a rearrangement is detected, the simulation is duplicated; one copy proceeds with forward shear, and the other with reverse shear. Each new state may thus lead to two more, and so on, revealing a catalog of states and rearrangements within some limits of deformation---represented as a transition graph in Fig.~\ref{fig:anneal}d \cite{Mungan19} that resembles the minimal graph in Fig.~\ref{fig:1}b. 
These graphs feature prominent strongly-connected components (SCCs)---clusters of states wherein any state is reachable from any other. Within these pockets of reversible plasticity, one finds many limit cycles and approximate or perfect return-point memory~\cite{Regev21}. Relatively few rearrangements exit each SCC and lead irreversibly to a new one. From this perspective, mechanical annealing moves among SCCs and finds one that sustains a limit cycle of the chosen $\gamma_0$. The topology of these graphs is a meaningful way to characterize energy landscapes and annealing, and to formally test the memory behavior of materials and models~\cite{Mungan19b,Terzi20,Mungan19a,Ferrari22,Movsheva23}. Although the Preisach model has neither transients nor annealing, in Sec.~\ref{sec:interact} we will show that it can be a starting point for this discrete approach to annealing in solids.

\subsection{Annealing and memory capacity}\label{sec:annealcap}

In light of annealing, the memory readout of Fig.~\ref{fig:amplitude}c is no longer so simple. Is the rise in differences past the remembered strain explained by the Preisach model, or by new rearrangements that had not been activated by annealing? If an annealed material has return-point memory and can thus acquire a memory from a single cycle, was the memory in Fig.~\ref{fig:amplitude}c formed over many cycles, or just in the final cycle before readout? Testing with multiple memories offers a resolution~\cite{Fiocco14,Adhikari18,Mukherji19,Keim20,Keim22}. 
First, after the material is annealed with one amplitude (4\%), just one cycle of a smaller amplitude (3\%) writes a new memory (Fig.~\ref{fig:anneal}e). The subsequent ``C~4,3'' readout in Fig.~\ref{fig:anneal}f is consistent with the Preisach model within the annealing envelope, with a steep rise in rearrangements outside it. Second, a “ring-down” protocol attains reversible plasticity without imposing an envelope of annealing (Fig.~\ref{fig:anneal}c). The resulting ``RD~4,3'' readout has the same memory content, but many of the rearrangements past 4\% are missing. With this technique one can also test the order of memories (``RD~3,4'') or omit writing altogether (``RD''). The simplest case of constant-amplitude shear in Figure~\ref{fig:amplitude}c thus blends two kinds of memory: one that must be ``trained'' over many cycles, and one that is merely written.

\section{GLASSY MEMORIES AND GLASSY MATTER}
\label{sec:interact}

\begin{figure}[b]
    \includegraphics[width=6.3in]{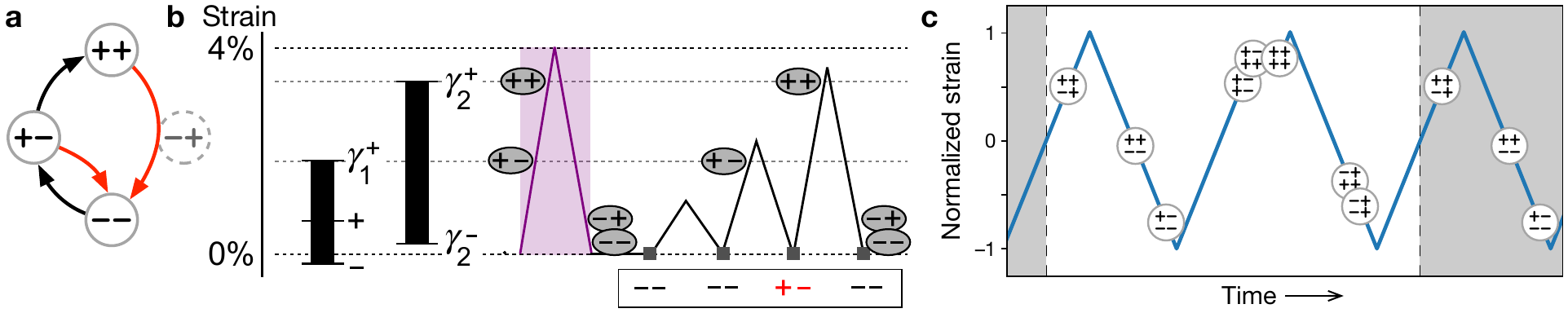}
    \caption{Simple behaviors of interacting hysterons.
    \textbf{(a)} Transition graph with nodes showing the states of two hysterons with a cooperative interaction. Red and black edges correspond to forward and reverse shear. The rightmost edge is an avalanche in which hysteron 1 switching to ``$-$'' forces its partner to immediately do the same~\cite{Mungan19}.
    \textbf{(b)} Two frustrated hysterons can store an amplitude of asymmetric shear~\cite{Lindeman23preprint}. Thresholds are represented at left; for clarity, only the crucial dependence of $\gamma^-_1$ on hysteron 2 is shown. At right, a minimal strain protocol shows that the hysterons' states upon returning to $\gamma = 0$ depend on amplitude, violating return-point memory. 
    \textbf{(c)} Multiperiodic orbit. Due to frustration, 4 hysterons return to the same state only after 2 periods of driving. 
    Panel \emph{b} courtesy of Chloe Lindeman. 
    Panel \emph{c} adapted from Ref.~\citealp{Keim21} (CC BY).
    }
    \label{fig:interact}
    \end{figure}

While the Preisach model explains some remarkable memory behaviors, it assumes that rearrangements do not interact. In fact, the extended displacement field around one rearrangement (Fig.~\ref{fig:amplitude}d), required by boundary conditions~\cite{Eshelby57,Nicolas18}, may be coupled to another nearby. Thus one rearrangement can facilitate or even trigger a like transformation nearby via a ``cooperative'' or ``ferromagnetic'' interaction. In the minimal transition graph of Fig.~\ref{fig:interact}a, two hysterons flip separately during forward shear (black arrows) but together during reverse shear, not because they have the exact same $\gamma^-$, but because the first destabilizes the second. This second flip might trigger further events in its own adjacent regions, and so on---a mechanism for avalanches that are a hallmark of slowly-deformed amorphous solids~\cite{Nicolas18,Jerolmack19,Antonaglia14,Bonn04,Denisov16}. Nonetheless, memory is qualitatively unchanged by these interactions: Sethna \emph{et al.}~\cite{Sethna93} proved that a system with cooperative interactions still has a fixed order of transitions (``no-passing'' \cite{Middleton92}) and hence return-point memory.

One rearrangement may also stabilize another nearby soft spot, so that e.g.\ a greater imposed strain is required to cause another rearrangement. The quadrupolar pattern of Fig.~\ref{fig:amplitude}d implies this kind of ``frustrated'' or ``antiferromagnetic'' interaction in which a pair of similar rearrangements can compress the elastic material between them; more conclusive evidence comes from analyses of molecular dynamics simulations~\cite{Mungan19,Szulc22}. Frustration is associated with ``glassy'' physics since it can give a system many mutually exclusive ways to relax, almost all of which stay far from a global energy minimum~\cite{Binder86}.

With frustration, return-point memory is no longer assured~\cite{Sethna93,Hovorka08}. Nonetheless, several studies of hysteron models have shown that instead of destroying return-point memory, frustration seems to degrade it only slightly---and in exchange, creates a wealth of new memory behaviors that we can predict, understand, or design. In these models the thresholds 
$\gamma_i^\pm$ from Sec.~\ref{sec:rpm} are functions of the other hysterons' states. The simplest models use a perturbative coupling matrix $J_{ij}$:
\begin{equation}
\gamma_i^\pm = (\gamma_i^\pm)_0 - \sum_{j \neq i} J_{ij} S_j
\label{eqn:interacting}
\end{equation}
where $(\gamma_i^\pm)_0$ is the unperturbed threshold, and the $j$th soft spot can be in state $S_j = \pm 1$. $J_{ij} < 0$ corresponds to a frustrated interaction. 

Lindeman \emph{et al.\ }have shown that just two frustrated hysterons can have a new memory behavior~\cite{Lindeman23preprint}. Figure~\ref{fig:interact}b shows the crucial feature of this system: how the threshold $\gamma_1^-$ depends on the state of hysteron~2. 
This pair is driven with a protocol similar to the memory tests of Fig.~\ref{fig:amplitude}f, but here the driving is asymmetric, with $\gamma \ge 0$---perhaps corresponding to the loading of a bridge by a series of trucks with different weights. Since each cycle ends at the bounding turning point $\gamma = 0$, a system with return-point memory would end every cycle in the same state, and attempting to write a smaller memory after a larger one (analogous to Fig.~\ref{fig:amplitude}f and Fig.~\ref{fig:anneal}e) would fail. Instead, the state of hysteron 1 upon returning to $\gamma = 0$ depends on the preceding amplitude---it ``latches'' in the ``$+$'' state for a portion of readout, and cannot return to ``$-$'' until a larger $\gamma_\text{read}$, approaching the remembered amplitude, has switched hysteron 2.
Changing state upon a reduction in amplitude, and changing back when the larger amplitude is resumed, is the essential behavior for storing multiple amplitudes. For symmetric driving this can be accomplished with non-interacting hysterons, but for asymmetric driving frustration is the key. 

Adding just one more hysteron with frustrated interactions leads to a staggering array of possibilities, surveyed by van Hecke~\cite{Hecke21}---even in the response to constant-amplitude driving. One new behavior resembles a striking phenomenon observed in molecular dynamics simulations of amorphous solids: the steady-state response is sometimes ``multiperiodic'' or ``subharmonic,'' with a period of particle motions that is an integer multiple of the period of driving---as in the action of a retractable pen or a mechanical counter. This defies the expectation for return-point memory that repeating a cycle of deformation should have the same result as performing it once. 
This behavior requires frustrated interactions, but it can arise in groups of just 3 or more hysterons. Much more common are period-1 (i.e. non-multiperiodic) orbits preceded by transients of 2 or more cycles---also impossible without frustration~\cite{Keim21,Lindeman21}. 
Together with evidence from spin glasses~\cite{Deutsch03}, these behaviors show that frustrated systems can count how many times a deformation was repeated. 
In a further expansion of memory capacity, Lindeman and Nagel~\cite{Lindeman21} found that while return-point memory requires that a memory at one amplitude be erased by driving with a larger one, with frustration the readout of the new memory contains a trace of the old one.

\begin{marginnote}[]
    \entry{multiperiodic}{Behavior by which a material returns to a previous state only after multiple cycles. Also termed ``subharmonic.''}
\end{marginnote}

\subsection{Frustration in glassy matter}

How close are interacting hysterons to the physics of amorphous solids? The molecular dynamics simulations of Lavrentovich \emph{et al.}~\cite{Lavrentovich17} found that within large ensembles of random systems, the occurrence of multiperiodic orbits decays exponentially with their period. A simple interpretation of this result is that as driving explores a system's states, there is a constant probability $< 1$ of revisiting a state and closing the orbit. Remarkably, ensembles of many random hysteron systems, with as few as 3 hysterons each, show this same behavior~\cite{Lindeman21,Keim21}---suggesting a connection rooted in the essential physics of glassiness. 
More directly, Szulc, Mungan, and Regev~\cite{Szulc22} carefully teased apart the mechanisms for multiperiodicity in a molecular dynamics simulation, and confirmed that frustrated interactions among soft spots are essential. However, their results favor a more sophisticated approach to modeling than Eq.~\ref{eqn:interacting}: interactions perturb $\gamma_i^+$ and $\gamma_i^-$ independently via coupling matrices $A^+_{ij}$ and $A^-_{ij}$, which can temporarily invert a hysteron's thresholds ($\gamma^+_i < \gamma^-_i$), putting it in a single state that cannot be switched.

One more step toward faithful mesoscopic models is to compute interactions among rearrangements according to their relative positions: soft spots aligned along the direction of shear, or perpendicular to it, tend to interact cooperatively; those aligned at $\pm 45^\circ$ to these directions tend to be frustrated~\cite{Nicolas18}.
This fact is crucial in models that include a material's spatial organization~\cite{Nicolas18,Keim21}. Close to interacting hysteron models is the deterministic ``integer automaton'' model~\cite{Talamali12,Khirallah21}: any cell in a square lattice can rearrange, and each rearrangement alters the stress field at nearby cells, using a discrete approximation of the theoretical result~\cite{Eshelby57}. The cells are not hysterons with two states, but rather integers that can be incremented or decremented indefinitely. While the model can mimic frustrated hysterons under cyclic driving at small amplitudes, it also shows extended transients and a transition to irreversibility at larger amplitudes---difficult or impossible behaviors for hysterons~\cite{Khirallah21}. This model's transition graphs have statistics similar to those from molecular dynamics simulations~\cite{Kumar22}.

The forms of glassy memory capacity we have discussed seem comprehensible only in situations where return-point memory predicts that returning a system to a previous strain would restore it to a previous state, but instead one observes a difference. In experiments, one might use this idea to isolate regions with frustrated dynamics for closer study, providing new ways to study glassy physics up close.

\section{HYSTERESIS AND MEMORY IN CRUMPLED SHEETS AND MECHANICAL METAMATERIALS}

\subsection{Connecting amorphous solids to crumpled sheets}

As we have seen, amorphous solids inspire models of coupled hysterons with which one can seek these materials' most exotic or surprising possible behaviors.
But this same abstraction has helped researchers pursue memory in many more physical systems. 

\begin{figure}[tb]
\includegraphics[width=5.25in]{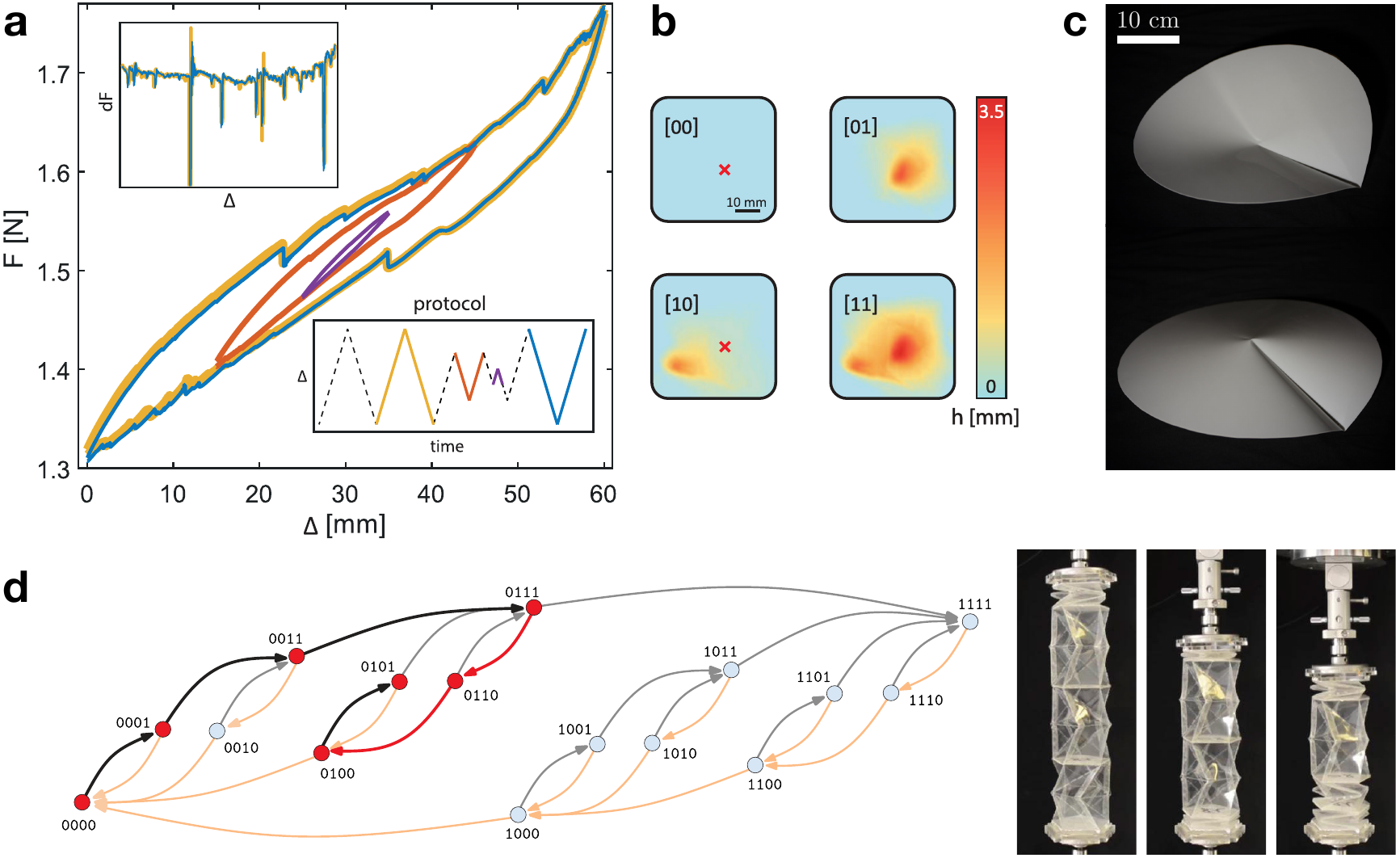}
\caption{
Memory from hysterons in crumpled and folded sheets. 
(\textbf{a}) Force-versus-displacement curves as a pre-crumpled sheet is cyclically compressed from two ends. 
After tracing two nested sub-cycles, the system can reproduce the outermost hysteresis loop, including much of the detailed structure of the force drops. (\textbf{b}) Height differences in a top-down view of a portion of the sheet, as various elements of the sheet ``snap-through'' between different stable configurations. 
(\textbf{c}) Bistability of a circular sheet with a single fold. 
(\textbf{d}) Origami bellows consisting of four bistable units, tracing out one possible path in its $16$-state transition graph. 
Panels \textit{a} and \textit{b} reproduced with permission from Ref.~\citealp{Shohat22} (CC BY-NC-ND 4.0). 
Panel \textit{c} adapted with permission from Ref.~\citealp{Lechenault15}: F.\ Lechenault and M.\ Adda-Bedia, \textit{Phys.\ Rev.\ Lett.} 115, 235501 (2015). Copyright 2015 by the American Physical Society. 
Panel \textit{d} adapted with permission from Ref.~\citealp{Jules22} (CC BY 4.0).
}
\label{fig:crumple}
\end{figure}

As a first example, consider the mechanical response of a cyclically-crumpled sheet~\cite{Ben-Amar97,Matan02,Witten07,Cambou15,Gottesman18,Croll19,Fokker19,Timounay20,Dawadi24}.
Figure~\ref{fig:crumple}a shows a force versus displacement curve from a thin plastic sheet that has first been prepared by crumpling it and then unfurling it into an approximately flat configuration, and then cyclically compressing it~\cite{Shohat22}. %by a time-varying displacement $\Delta$
The curves display nested hysteresis loops, with a series of intermittent force drops throughout. 
The driving protocol is shown in the inset: A large driving cycle is applied first, followed by two smaller cycles and a final large cycle. 
Remarkably, the final cycle produces a nearly-identical hysteresis loop as the first, even reproducing the dominant force drops. 
These are hallmarks of return-point memory. 
Shohat, Hexner, \& Lahini~\cite{Shohat22} were able to find the hysterons responsible for this macroscopic response: the sheet  mesoscopic regions that ``snap through'' between different stable configurations, pictured in Fig.~\ref{fig:crumple}b. 
Although return-point memory does not require any coupling between the hysterons, Ref.~\citealp{Shohat22} teased apart the coupling between pairs of hysterons in their sheet, showing that their interactions can be cooperative or frustrated. 
In a subsequent study, Shohat \& Lahini~\cite{Shohat23a} identified excess energy dissipation as a harbinger of memory formation in crumpled sheets, forming another parallel between memories in crumpled sheets and in disordered particulate matter.

\subsection{Origami hysterons}

One way to elucidate the memory in a randomly-crumpled sheet is to study hysterons in sheets with just a few folds. 
Perhaps the simplest case is shown in Fig.~\ref{fig:crumple}c: 
when a point lying on a crease is pushed normal to the sheet, it can snap through to another stable configuration~\cite{Hanna14}. Adding folds~\cite{Lechenault15} or changing the shape of the boundary can control the switching thresholds~\cite{Waitukaitis15} of this kind of hysteron. 
Intriguingly, bistability survives even when a hole is cut through the sheet along the fold \cite{Yu22}, removing the point-like elastic singularity. Changing the size and shape of the hole offers further control of the switching thresholds.

There is another way to achieve bistability: 
some origami have folding modes that traverse an energy barrier~\cite{Greenberg11,Silverberg15,Reid17,Liu18}. 
By connecting several such units together, one can build up metamaterials with many internal states and a set of allowed transitions between them. 
The bellows in Fig.~\ref{fig:crumple}d shows a design that possesses memory of its previous deformation~\cite{Jules22}. 
Each of its four modules is based on the same folding pattern, but with slightly different angles so that each module expands and collapses at a different force  $F^\pm_i$. Labeling the deployed state as $1$ and the collapsed state as $0$, one may thus list the states of the system with four-bit strings such as $0110$. 
Any of the $2^4$ possible states of this system can be reached through the appropriate series of compressions and expansions, as shown by the transition graph in Fig.~\ref{fig:crumple}d; the bellows has return-point memory. 
A similar mechanical behavior can be realized in a flexible truss structure \cite{Yasuda17}, which resembles the origami bellows with faces removed, and edges replaced by springs. 

\begin{figure}[b]
\includegraphics[width=5.0in]{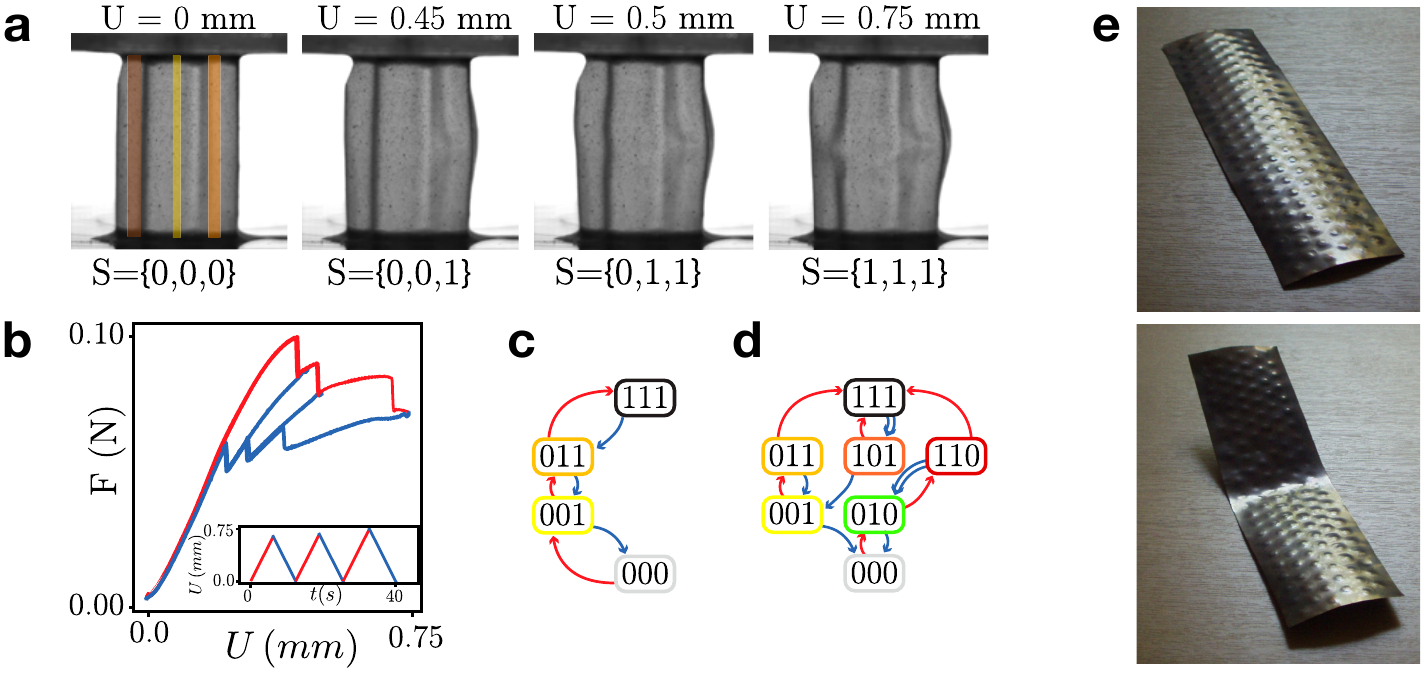}
\caption{
Coupled hysterons in mechanical metamaterials. 
(\textbf{a-d}) Corrugated silicone sheet with three ridges that act as hysterons upon vertical compression by a distance $U$.
(\textbf{b}) The mechanical response of the system shows sharp drops and recoveries of the force upon loading and unloading to different amplitudes, where each hysteron transitions between its states. 
(\textbf{c}) Transition graph corresponding to panels (\textit{a}) and (\textit{b}). 
(\textbf{d}) The system experiences a proliferation of new states and transitions if the top plate is tilted by a small amount. 
(\textbf{e}) A metal sheet with a lattice of bistable dimples can access a multitude of mechanically stable configurations; two of them shown here. 
Panels \textit{a-d} reproduced with permission from Ref.~\citealp{Bense21} (CC BY 4.0). 
Panel \textit{e} reproduced with permission from Ref.~\citealp{Seffen06}. 
%reproduced with permission from Reference~\citealp{Seffen06}.
%%% Need permissions from Elsevier before ARCMP publishes. %%%
}
\label{fig:Hecke}
\end{figure}

\subsection{Coupled hysterons in mechanical metamaterials}

The simple hierarchical structure of Fig.~\ref{fig:crumple}d suggests that interactions between hysterons in this system are negligible \cite{Jules22}. 
This state of affairs may be desirable for straightforward control of a system's state, as in the Preisach model. But as with amorphous solids in Sec.~\ref{sec:interact}, interactions make many more behaviors possible. Metamaterials are an opportunity to tailor these interactions and design mechanical memories and computations at the limits of glassy matter~\cite{Merkle93, Treml18, Yasuda21}. 

To demonstrate tunable interactions, Bense and van Hecke \cite{Bense21} made a corrugated silicone sheet in the shape of a cylindrical arc with three sinusoidal ridges (Fig.~\ref{fig:Hecke}a). As the sheet is compressed and released, each ridge acts as a hysteron that buckles and straightens, causing the jumps in the three nested force-versus-displacement curves of Fig.~\ref{fig:Hecke}b. Initially, just four states are reachable, shown in Fig.~\ref{fig:Hecke}c with their transitions. 
However, in this design the switching thresholds---and hence the effects of interactions---can be tuned via the boundaries. 
After tilting the bottom plate by a small angle ($\mathcal{O}(10)$ milliradians), one obtains the dramatically different transition graph in Fig.~\ref{fig:Hecke}d. 
More states are reachable, and an instance of ``scrambling'' is observed, shown by the double-lined blue arrows.
Comparing the down transitions from the states 111 and 110, one may probe the influence of the third hysteron (in the 1 or 0 state, respectively) on the first two hysterons (which both begin as 1 in this pair of states). 
The effect is that the second hysteron switches first when the third hysteron is `on' ($111 \rightarrow 101$), but the first hysteron switches first when the third hysteron is ‘off' ($110 \rightarrow 010$), making return-point memory impossible. 
This is an example where interactions are strong enough that their presence can be inferred from a transition graph.
Other material systems are being designed with strong interactions between bistable hysteretic elements~\cite{Muhaxheri24}, often with an eye towards performing logic operations \textit{in materia}~\cite{Merrigan22,Hyatt23}. 
One active avenue is the development of architectures that allow one to rationally target particular transition pathways~\cite{Liu24}.

\subsection{Elastic shape memory}

When the clamping is removed from the corrugated sheet in Fig.~\ref{fig:Hecke}a, the configuration of bistable ridges becomes coupled to the global deformations of the sheet~\cite{Meeussen23}. 
Such coupling also occurs in ``puckered sheets'' consisting of crystalline arrays of localized Gaussian curvature \cite{Seffen06, Faber20, Plummer22}. 
This coupling endows 
these materials with a kind of ``elastic'' shape memory \cite{Seffen06}, whereby large-scale bending or twisting of the sheet can change the state of the small-scale bistable elements, ``freezing'' an overall shape once formed. 
The same behavior can also emerge in disordered analogs \cite{Oppenheimer15}. 
In some cases, the detailed state of the puckers does not uniquely determine the global shape --- that is, there may be more than one stable macroscopic conformation for a given microstate of the bistable elements \cite{Liu23}. These examples are re-creations of a memory we may take for granted in malleable solids: a memory of shape ``written'' by plastic deformations and encoded in particle arrangements and bonds.

\subsection{Metamaterials that defy coupled-hysteron models}

Finally, some metamaterials with bistable elements can push beyond the limits of hysteron models by bringing in other physical interactions. Buckled beams in a material can resemble hysterons~\cite{Merrigan22}, but work by Ding, Kwakernaak, and van Hecke~\cite{Ding22,Kwakernaak23} 
describes systems where specially-designed contacts between beams are made, broken, and evolve under friction continuously in time. Although these systems show discrete memory behaviors of the kind described above, the kinematics and mechanisms of these behaviors are described poorly by discrete (hysteron-based) degrees of freedom. 

\section{MEMORY AND DEGREE OF DISORDER: FROM AMORPHOUS SOLIDS TO POLYCRYSTALS}

Many non-designed materials feature large crystalline domains unlike anything found in an amorphous solid. Indeed, the conventional starting point for constructing theories of solids has been to treat these materials as extensions of the tremendously successful theories of crystalline matter.
Thus it is \emph{a priori} unclear whether amorphous or crystalline matter is the better prototype for memory in polycrystals.

\begin{figure}[b]
\includegraphics[width=6.3in]{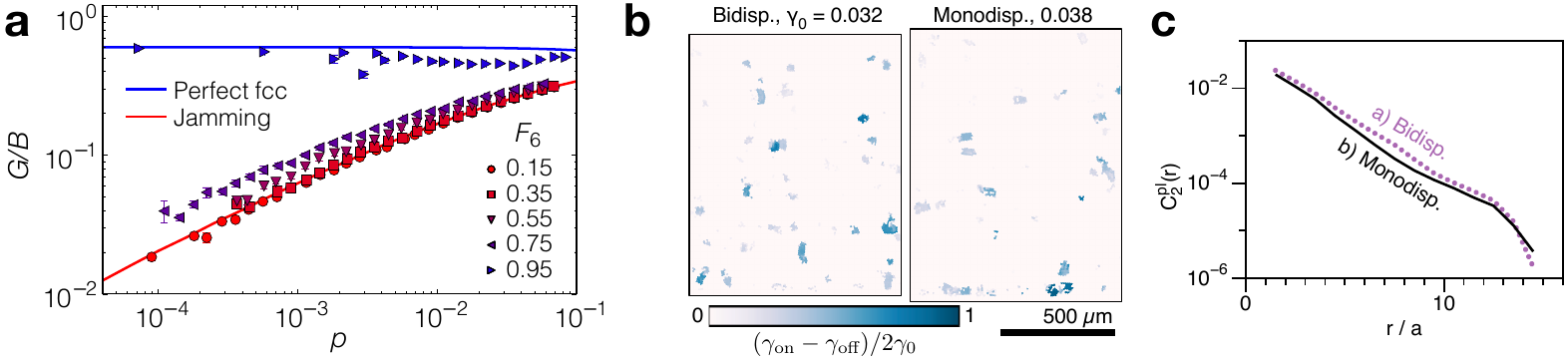}
\caption{
Bulk mechanics and mesoscopic rearrangements in polycrystals. 
(\textbf{a}) 
Ratio of shear modulus to bulk modulus for five systems that span the spectrum from totally disordered ($F_6 \approx 0$) to perfect crystal ($F_6 = 1$), where $F_6$ measures the correlations of bond-orientational order among neighboring particles. 
Four of the systems exhibit a vanishing ratio of shear to bulk modulus ($G/B$) in the limit of vanishing pressure, a hallmark of the jamming transition in amorphous solids. 
Only the system with $F = 0.95$ exhibits the plateau in $G/B$ that is characteristic of a crystal. 
(\textbf{b}) Comparison of rearranging regions in a more amorphous sample (left) and a more polycrystalline sample (right). Particles are colored according to the degree of hysteresis in their response to shear (thresholds $\gamma^+$ and $\gamma^-$ are labeled $\gamma_\text{on}$ and $\gamma_\text{off}$). The appearances and statistics of rearrangements are similar between sample types. 
(\textbf{c}) Two-point cluster function from these two systems, showing how the average correlation within a rearranging region decays with distance from the region's center.
The data from the amorphous solid and from the polycrystal track each other remarkably well. 
Panel \textit{a} reproduced with permission from Ref.~\citealp{Goodrich14}. 
%reproduced with permission from Reference~\citealp{Goodrich14}.
%%% Need permissions from Elsevier before ARCMP publishes. %%%
Panels \textit{b} and \textit{c} reproduced with permission from Ref.~\citealp{Keim15}.
}
\label{fig:crystal}
\end{figure}

To elucidate this issue, Goodrich \emph{et al.}~\cite{Goodrich14} produced a set of three-dimensional jammed systems along a spectrum from disordered to crystalline. 
They found that adding only a small amount of disorder to a crystal brings its behaviors closely in line with the most disordered case (Fig.~\ref{fig:crystal}a), consistent with other studies~\cite{Katgert09,Mari09}.
This view of polycrystals
extends to their plastic response: 
in simulations they exhibit localized rearrangements under loading \cite{Katgert09,Shiba10,Biswas13}, which arise in comparatively softer regions \cite{Rottler14}. 
In experiments, 
Keim and Arratia \cite{Keim15} created 2D jammed solids similar to the amorphous samples in Fig.~\ref{fig:amplitude}, but switched from bidisperse particle sizes to monodisperse, forming crystalline regions separated by disordered grain boundaries.
Remarkably, localized rearrangements are nearly the same diameter and magnitude whether the sample is amorphous or polycrystalline (Fig.~\ref{fig:crystal}b,c); the main difference is simply that the polycrystal has fewer of them. 
These rearrangements are predominantly at grain boundaries \cite{Keim15,Sharp18}, although particles within crystalline regions can rearrange as well \cite{Teich21}. 
These microscopic results suggest that polycrystals can form memories of cyclic driving. 
Indeed, this is in line with experiments showing memories in a bubble raft in an annular geometry, where one can see crystalline domains \cite{Mukherji19}. How the strength of the memory signal, or the capacity for multiple memories, varies with the degree of disorder in polycrystals remain open avenues of study. 
For instance, amorphous solids appear to present strong memories of cyclic driving near yielding \cite{Mukherji19}, whereas 2D Lennard-Jones polycrystals can coarsen into monocrystals as yielding is approached \cite{Jana18,Khushika23}---raising the possibility that near yielding, polycrystals can lose their memory capacity altogether. 

Intriguingly, while the crystalline domains in a polycrystal seem to have few of the rearrangements that might encode memories of amplitude, they do store at least one other kind of memory:
the direction of the last large shear when a sample was prepared. 
Teich \emph{et al.}~\cite{Teich21} analyzed experimental data from 2D jammed amorphous bidisperse particle systems \cite{Keim14,Keim15}. 
They measured slight anisotropies at the particle level, similar to the amorphous solid results in Fig.~\ref{fig:anneal}(a, b)---but here, the memory of direction is strongest within crystalline domains, where it corresponds to a slight distortion of the lattice \cite{Teich21}. This result shows that crystallinity and mechanical memory are not mutually exclusive.

\section{EMERGING AND FUTURE DIRECTIONS}

\subsection{Dynamics}

Particle rearrangements are not instantaneous, due to inertial, viscous, or thermal considerations~\cite{Nicolas18,Falk11,Greer13,Ness22}---nor are these timescales uniform in a disordered system. Away from the quasistatic limit, some relaxations may not keep up with a deformation, leading to nonlinear wave propagation~\cite{McCall1994,Guyer1999} and to new possibilities for memory: the influence of a ``slow'' rearrangement on its neighbors will change when the strain rate is increased, or if a turning point of a cyclic deformation is reached too soon~\cite{Kern04,Lindeman23}, so that when the material comes to rest it encodes past dynamics. Like extended transients and multiperiodicity, this sensitivity may even play out over many cycles~\cite{Lindeman23}. 
One exciting possibility is that a suitable readout protocol could recover the rates of past deformations, a task that today is usually relegated to failure analysis~\cite{Greer13,Jerolmack19}. 
Some of these ideas may also apply to mechanical metamaterials, where it was recently shown that driving the material dynamically can allow one to reach states that would be unattainable using quasistatic deformation \cite{Jules23}. 

\subsection{Thermal or mechanical noise}

One might suppose that noise weakens memories gradually over time, as in dilute suspensions~\cite{Paulsen14}. A different picture emerges from recent molecular dynamics simulations of amorphous solids by Majumdar and Regev~\cite{Majumdar23}, showing that even if the material's temperature is too low to cause rearrangements in the material at rest, it may cause the thresholds $\gamma^\pm_i$ to fluctuate during deformation. These fluctuations may have no net effect in any one cycle of shear, but after a finite number of cycles, the sequence of rearrangements may eventually be scrambled---a change that, via interactions, can disrupt the limit cycle. The steady-state behavior is thus a series of limit cycles punctuated by transients. Deforming at finite frequency seems to heighten this effect. Noise may thus be a tunable way of erasing memories without forming new ones. 

\subsection{Particle interactions}

The above studies of memory in amorphous solids focus on frictionless spherical particles with short-range repulsion, leading one to ask how to generalize their results. Fortunately, the many kinds of amorphous solids and polycrystals tend to be more alike than different in their physics~\cite{Cubuk17}---for example, most molecular dynamics simulations use Lennard-Jones potentials that have short-range attraction, but they agree qualitatively with experiments featuring long-range repulsion~\cite{Keim20}. Nonetheless, the study of memory is already finding surprises as the details of interparticle interactions are varied.

\subsubsection{Friction}

Most solid grains exhibit static friction: contacts sustain tangential forces as well as normal ones. Considering non-spherical particles leads to similar complications. These additional mechanical constraints underpin the rigidity of matter as consequential as soil and km-scale asteroids~\cite{Jerolmack19,Daniels13,Walsh18}, and they dramatically expand the set of stable configurations of a fixed number of particles.
% For $N$ spheres in a 3D packing, one might interpret friction as increasing the degrees of freedom that are coupled to driving but cannot relax, from $\sim$$3N$ (position) to at least $5N$ (position and orientation). 
Do these new possibilities simply enhance the mechanical memory behaviors we have already discussed?

One hint comes from deforming a packing with constant amplitude. In experiments, the arrangement of particles can reach a limit cycle even as the contact forces fluctuate indefinitely---the coupling of the system's state to the boundaries is qualitatively distinct~\cite{Kollmer19,Slotterback12}. Indeed, simulations and experiments by Benson \emph{et al.}~\cite{Benson21} showed that the strength and form of a memory readout differed between measurements of translation and rotation. More striking, reducing amplitude seemed to erase memory, preventing the nesting of multiple memories allowed by return-point behavior---a seeming reduction in memory capacity due to friction.

While memories of amplitude in these everyday materials remain a puzzle, friction enables a different paradigm for memory. Until it slips, each contact preserves the relative positions of two surfaces when they first came together, even as the packing and the particles are deformed further. Simulations by Candela~\cite{Candela23} slowly compress a packing to form new contacts, while applying a varying shear strain that is orthogonal to the compression. As the packing is then returned to zero shear strain and decompressed, the contacts slip in reverse order, each one changing the shear stress according to the shear strain at its formation---``playing back'' the time-reversed waveform of shear strain. The hierarchy and time-reversal inherent in these frictional contacts is reminiscent of how nested return-point memories of amplitude must be written and read in reverse order---but here, storing memory requires avoiding the particle rearrangements that encoded memories in that earlier example.

\subsubsection{Tenuous solids: Memories in gels}

A gel is a solid with a much more tenuous structure: its constituent particles are held together by attractive contacts, leaving voids that are filled by a fluid. 
To what extent do memory effects occur in these materials?
Schwen \textit{et al.}~\cite{Schwen20} applied hundreds of shear cycles to a colloidal gel, followed by readout cycles. 
In confocal images the particle trajectories were markedly more irreversible beyond the training strain, indicating a memory. 
The memory could also be read out by shearing in a direction \emph{orthogonal} to the training. 
Analysis of the packing structure showed that the more-reversible configurations have a comparatively larger number of nearest-neighbor contacts. 
Here, the bond structure appears to be the salient aspect of the system for memory formation, in contrast to the localized rearrangements that were so important in other amorphous solids. 

Chattopadhyay and Majumdar \cite{Chattopadhyay22} observed similar memories in a dense suspension of cornstarch particles in oil. They were able to read out two trained shear amplitudes, so long as the smaller amplitude was applied last. 
They propose a microscopic picture where the material forms loose strands \cite{Doorn18} that can become taut and break under shear. 
Training develops a population of strands that break just past the training strain, analogous to how mechanical memories form in cross-linked biopolymer networks \cite{Schmoller10}.

The idea that a system ``learns'' to be marginally stable with respect to a repeated drive is a powerful concept that links these memories in gels with the memory of annealing in Sec.~\ref{sec:anneal} and even with memories in non-Brownian suspensions and charge-density waves \cite{Keim19}.
Nevertheless, gels have other memory behaviors that seem to arise specifically from the way their structure is remodeled at rest \cite{Chen23}.

\subsection{Training function into materials}

Pondering the variety of memories that have been observed or predicted, one starts to wonder what other kinds of annealing, training, and memory are possible, especially with driving beyond uniaxial shear. Discovering the answers is more than an academic exercise: If a material is led to a well-defined state that encodes memory of an input, there may be other well-defined properties of the material that come along for the ride. Among the earlier examples, memories of a direction or strain amplitude alter the mechanical response as a function of direction or strain---storing information has become a strategy for tuning mechanical properties within some design space. 
This idea is appealing because training may be much easier than controlling bulk properties by specifying microstructure, it can be done \emph{in situ} as needs change, and it can resemble the desired response---for example, a material is sheared to modify its response to further shear.

A distinction is apparent in this review between ``short-term'' memory that may be formed or erased by as little as one cycle of deformation, by selecting a memory-encoding state from an existing landscape of possibilities,
and ``long-term'' memory that is formed by gradually annealing or remodeling the available states of the system toward some purpose. The latter is exemplified by ``physical learning'' in transport or mechanical networks \cite{Pashine19,Stern23,Anisetti23} 
in which the material develops a global coordinated response to specific inputs, similar to an associative memory trained over many iterations in an artificial neural network, rather than a short-term buffer. If we suppose that short-term memories like the ones we describe are also desirable for systems that compute and adapt, can those capacities be trained as well?

\section{CONCLUSION}

While we have only touched on a fraction of the fascinating memory behaviors of solids, several themes are apparent:

\begin{itemize}
    \item \textbf{Solids form memories.} Memories are among the solid phase's best features. Because a solid at rest does not relax on experimental timescales, anything that changes a solid can form a memory. Amorphous solids are just one instructive example---once we understand some of the fascinating memory behaviors that arise from these materials' mixture of localization, disorder, hysteresis, and interactions, we can observe or design those same ingredients in many more systems.
    \item \textbf{Studying memory is studying memory capacity.} Once readable information is identified, there is much more to be learned by testing the limits of what can be preserved, and what must be forgotten. Even among memories of shear strain, it is the details of capacity---its dependence on the sequence of writing and reading, the polarity of driving, the system size, glassiness---that signify important features of these systems' physics, and lead to distinctions and commonalities among examples. 
    Even when models fall short, considering capacity gives us more specific and varied ways to probe solid matter, and language to describe the results.
    \item \textbf{Hysterons are more than just hysteresis.} Hysterons were  proposed as fitting parameters for hysteresis, but in solids they correspond to the mesoscopic relaxations that control plasticity. Remarkably, these models capture essential features of behaviors that are completely unlike hysteresis---for example, counting. In amorphous solids and crumpled sheets, this simplified framework lets us focus on the collective physics of interacting relaxations, while in mechanical metamaterials it shows how memories and computations may be built up from simple elements. Further afield, the idea of training hysterons' thresholds and interactions gives us a specific route to training function.
    \item \textbf{Memory navigates an energy landscape.} The disorder of an amorphous solid can lead us to see it as the embodiment of arbitrary happenstance among a sea of possibilities---so that once a unique configuration of particles has been lost, one should never expect to see it again. Yet the orderly memory behaviors we have described prove that this kind of return is commonplace: that up to strains of several percent, a finite and highly connected set of states is sure to emerge---and by understanding how to return to a past state, we learn to store memories and even perform simple computations. The surprising success of this approach in experiments and simulations suggests that memory belongs with the fundamental study of glassy matter.
    \item \textbf{Memories bridge scales.} The work in this field has primarily focused on model soft matter systems in which writing and reading are all under a researcher's control. Yet the materials and scenarios for memory can scale far beyond these examples. To cite a few extremes, one may consider the mechanical memories that record history and predict response in earthquake fault zones---samples that are tens of meters thick and subject to varying GPa-scale stresses over tens of thousands of years~\cite{Jerolmack19}. Or, one might look to aging and failure of nanoscale structural components made of bulk metallic glass \cite{Cubuk17,Greer13,Adhikari23}. 
    In the world of artificial matter, 
    origami designs have long been recognized as scalable from the benchtop to, e.g., deployable space structures; the more recent concept of memory in these materials could broaden their capabilities. 
    \end{itemize}

The study of memory continues to reveal new possibilities.
As researchers race to create materials that adapt and respond intelligently to their mechanical environments, the question of memory is central \cite{Stern23}.
Studying memory in amorphous solids has  shown how some complex memory behaviors like short-term memory, which can emerge in living and artificial neural networks, can also be achieved by a different mechanism in a solid. 
Perhaps closest at hand are ways that living organisms compare stimuli across time, which are not much more complex than behaviors we have described and explained here~\cite{Hachen21,Riviere23}. 
Just as research on memory is teaching us more about solid matter and its dependence on the past, it is giving us new ways to shape the future.

\section*{DISCLOSURE STATEMENT}
The authors are not aware of any affiliations, memberships, funding, or financial holdings that
might be perceived as affecting the objectivity of this review. 

\section*{ACKNOWLEDGMENTS}
We are grateful to the many people who have shaped our understanding and appreciation of memory over years of interactions, several of whom also provided feedback on a draft of this article. A partial list must include Omri Barak, Karen Daniels, Mathew Diamond, Larry Galloway, Varda Hagh, Yoav Lahini, Chloe Lindeman, Craig Maloney, Enzo Marinari, Alan Middleton, Muhittin Mungan, Sidney Nagel, Abigail Plummer, Ido Regev, Srikanth Sastry, Martin van Hecke, Damien Vandembroucq, Zhicheng Wang, Zorana Zeravcic, and organizers and participants of the May 2023 CECAM workshop ``Pathways, Memory, and Emergent Computation in Nonequilibrium Systems,'' Sde Boqer, Israel. 
NCK was supported in part by a Research Grant from the Human Frontier Science Program (Ref.-No: RGP0017/2021).

%\bibliography{mem}

\end{document}